\documentclass[aps,showpacs,amsmath,amssymb,preprint,prd,%
floatfix,superscriptaddress]{revtex4}

\usepackage{graphicx}
\usepackage{amssymb}
\usepackage{amsmath}
\usepackage{dcolumn}
\usepackage{docs}%
\usepackage{bm}
\def\lsim{\mathrel{\rlap{\lower4pt\hbox{\hskip1pt$\sim$}}
    \raise1pt\hbox{$<$}}}               
\def\gsim{\mathrel{\rlap{\lower4pt\hbox{\hskip1pt$\sim$}}
    \raise1pt\hbox{$>$}}} 

\begin{document}

\title{Medium effects and the shear viscosity of the dilute Fermi gas 
away from the conformal limit}

\author{M.~Bluhm and T.~Sch\"afer}
\affiliation{Department of Physics, North Carolina State University, 
Raleigh, North Carolina 27695, USA}


\begin{abstract}
 We study the shear viscosity of a dilute Fermi gas as a function of the 
scattering length in the vicinity of the unitarity limit. The calculation
is based on kinetic theory, which provides a systematic approach to 
transport properties in the limit in which the fugacity $z=n\lambda^3/2$
is small. Here, $n$ is the density of the gas and $\lambda$ is the 
thermal wave length of the fermions. At leading order in the fugacity 
expansion the shear viscosity is independent of density, and the minimum 
shear viscosity is achieved at unitarity. At the next order medium effects 
modify the scattering amplitude as well as the quasi-particle energy 
and velocity. We show that these effects 
shift the minimum of the 
shear viscosity to the Bose-Einstein condensation (BEC) side of the 
resonance, in agreement with the result of recent experiments. 
\end{abstract}

\maketitle

\section{Introduction \label{sec:1}}

 Cold fermionic gases provide a unique arena for the study of strongly 
correlated matter. In these systems the $s$-wave interaction between 
atoms can be controlled by altering an external magnetic field. The 
dimensionless parameter that governs the interaction is $k_F a$, where 
$a$ is the $s$-wave scattering length and $k_F$ is the magnitude of 
the Fermi momentum. In a homogeneous Fermi gas the latter is related 
to the particle density by $n=k_F^3/(3\pi^2)$. The Fermi momentum also 
defines a temperature scale, the Fermi temperature by $T_F=k_F^2/(2m)$. 
Tuning the system into a Feshbach resonant state corresponds to the 
limit $k_Fa\to\infty$. In this regime the interaction cross section 
is limited only by unitarity. Equilibrium and non-equilibrium properties
of the dilute Fermi gas at unitarity have been investigated in a number
of experiments, for example~\cite{O'Hara:2002zz,Bartenstein:2004zz,Kinast:2005zz,Altmeyer:2007zz}. 

 In the unitarity limit the dilute Fermi gas is a scale and conformally 
invariant non-relativistic many-body system. At high temperature $T$ it 
behaves as a weakly interacting gas, but in the low $T$ limit it is a
strongly correlated quantum fluid, which shares many interesting properties
with other strongly interacting systems. An important example is nearly 
perfect fluidity, which has also been observed in the relativistic 
quark-gluon plasma~\cite{Schafer:2009dj,Adams:2012th}. 

 Detuning the system away from unitarity, scale invariance is lost. At low $T$ 
and positive $a$ a Bose-Einstein condensate (BEC) of strongly bound diatomic 
molecules is formed~\cite{Greiner:2003zz,Zwierlein:2003zz,Jochim:2003sci}. 
On the atomic side of the resonance, i.e.~for negative $a$, a 
Bardeen-Cooper-Schrieffer (BCS) superfluid state is realized at low $T$.
The crossover between the BEC and BCS regimes is known to be smooth
\cite{Bartenstein:2004zza,Bourdel:2004zz}. As the temperature is increased
the superfluid Fermi gas undergoes a phase transition to a normal fluid.
On the BEC side this is the Einstein transition, which occurs at a critical
temperature $T_c\sim T_F$. In the BCS regime the critical temperature is 
exponentially small compared to $T_F$. The maximum transition temperature 
$T_c/T_F$ is achieved slightly on the BEC side of the BEC-BCS crossover. 

 It is natural to ask how transport properties of the fluid change along 
the BEC-BCS crossover. The bulk viscosity, for example, vanishes at unitarity 
but is expected to be non-zero on either side of the Feshbach resonance. In 
kinetic theory bulk viscosity is thought to arise from scale invariance 
breaking encoded in the density dependence of the effective fermion mass 
\cite{Schaefer:2013oba}. The shear viscosity, on the other hand, is expected 
to be large in the BCS and BEC limits and become minimal close to unitarity. 
Indeed, kinetic theory in the high temperature limit predicts that the shear 
viscosity has a minimum exactly at unitarity \cite{Massignan:2005zz}. 
Experimental measurements of the shear viscosity at unitarity and at 
temperatures close to the phase transition have been reported in
\cite{Cao:2010wa,Cao:2011fh,Schafer:2007pr}. These experiments obtain values 
for $\eta/s$, the ratio of shear viscosity to entropy density, that are only 
a few times larger than the conjectured universal bound  $\eta\geq \hbar\,s
/(4\pi k_B)$ \cite{Kovtun:2004de}. Recent measurements~\cite{Elliott:2014nn} 
indicate, however, that the kinematic viscosity $\eta/n$ is minimized 
on the molecular side of the BEC-BCS crossover. 

 In this work we study the dependence of the shear viscosity on $k_Fa$ in 
kinetic theory. Kinetic theory can be viewed as a systematic expansion in 
the fugacity $z=n\lambda^3/2$ of the gas. Here, $n/2$ is the density per spin 
state and $\lambda=[(2\pi\hbar)/(mT)]^{1/2}$ is the thermal de Broglie wave 
length. At leading order in the fugacity the shear viscosity is independent 
of $z$ and has a minimum at $k_Fa\to\infty$ \cite{Massignan:2005zz}. This
minimum at unitarity is a simple consequence of the 
maximum in the vacuum cross section at resonance. We will show that 
including medium effects in the scattering amplitude shifts the minimum
away from unitarity. The physical reason for this behavior is related 
to Pauli blocking in the in-medium scattering amplitude, which is more 
efficient on the BCS side. Formally, the minimum in the shear viscosity 
arises from the competition between $(\lambda/a)^2$ and $z(\lambda/a)$ 
corrections to $\eta$. 
We will show that it is possible to compute all one- and two-body effects 
of order $\mathcal{O}(z(\lambda/a))$. In addition to in-medium corrections to the 
scattering amplitude, these terms arise from medium corrections to the 
quasi-particle energy and velocity. 

 Kinetic theory based on atomic degrees of freedom is reliable in the 
limits of high temperature, $T\gg T_F$, or weak coupling, $k_F|a|\ll 1$. 
Previous investigations have indicated that at unitarity kinetic theory 
is applicable at temperatures as low as $T/T_F\simeq 0.4$
\cite{Massignan:2005zz,Bruun:2008pra,Schaefer:2009px}. This condition
is satisfied for part of the data reported in \cite{Elliott:2014nn}.
Early studies of medium effects were reported 
in~\cite{Bruun:2005en,Bruun:2006kj}. Medium effects are also included 
in the $T$-matrix approaches of Enss et al.~\cite{Enss:2010qh} and 
Levin et al.~\cite{Guo:2010,He:2014xsa}.

 This paper is structured as follows: in Sect.~\ref{sec:2} we introduce 
a quasi-particle description for the dilute Fermi gas near unitarity. 
In Sect.~\ref{sec:31} we discuss the kinetic theory calculation of
the shear viscosity. A simple model based on medium-corrections to the 
cross section is described in Sect.~\ref{sec:32}, and a systematic 
expansion in powers of $z$ and $(\lambda/a)$ is given in Sect.~\ref{sec:33}.
We conclude in Sect.~\ref{sec:4}, and relegate details of the 
expansion to two appendices. 

\section{Quasi-particle description \label{sec:2}}

 In this section we introduce a quasi-particle model for the dilute 
Fermi gas near unitarity. The effective Lagrangian for non-relativistic 
spin 1/2 fermions interacting via a short range potential is 
\begin{equation}
\label{L_4f}
{\cal L} = \psi^\dagger \left( i\partial_0 + \frac{\nabla^2}{2m} \right) \psi 
 - \frac{C_0}{2} \left(\psi^\dagger \psi\right)^2 \, ,
\end{equation}
where the coupling $C_0$ is determined by the $s$-wave scattering length 
$a$. In the weak coupling limit we find $C_0 = 4\pi a/m$. In the high
temperature limit thermodynamic properties of the gas can be computed
as a systematic expansion in the fugacity $z$. This is the well known
virial expansion. The pressure is given by 
\begin{equation}
P= \frac{\nu T}{\lambda^3} \left( z+b_2z^2+\dots \right)\, , 
\end{equation}
with $\nu=2$ for two spin degrees of freedom. The second virial coefficient 
is obtained by summing the two-particle interaction to all orders. Near 
unitarity we get 
\begin{equation}
\label{equ:secondvirialcoefficient}
 b_2 = -\frac{1}{4\sqrt{2}}+\frac{1}{\sqrt{2}}
   \left(1+\frac{\sqrt{2}}{\pi}\frac{\lambda}{a}+\dots\right) \, ,
\end{equation}
which is valid on either side of the resonance. The temperature dependence 
of $b_2(T)$ is a measure for the scale invariance breaking. Given $P(\mu,T)$
we can compute other thermodynamic properties. The particle density $n=
(\partial P)/(\partial\mu)_T$ is given by 
\begin{equation}
 \label{equ:particledensity}
n = \frac{\nu}{\lambda^3}\left(z+2b_2z^2+\dots\right)\, ,
\end{equation}
and the entropy density $s=(\partial P)/(\partial T)_\mu$ is
\begin{equation}
 \label{equ:entropydensity}
s=\frac{5}{2} \frac{\nu}{\lambda^3}
   \left(z\left[1-\frac25 \frac{\mu}{T}\right]
   + b_2 z^2\left[1-\frac45\frac{\mu}{T}\right]
   + \frac25 T b_2'z^2+\dots\right) \, .
\end{equation}
We can construct a quasi-particle model consistent with these results 
by computing the fermion self-energy at order $z$. This corresponds
to summing the two-body interaction with a fermion in the heat bath 
to all orders. The fermion dispersion relation is given by $E_p=E_p^0+
\Delta E_p$ with $E_p^0=p^2/(2m)$ and $\Delta E_p={\rm Re}\,\Sigma(p)$, 
where $m$ is the mass parameter and $p$ is the magnitude of the momentum. 
The real part of the on-shell fermion self-energy near unitarity reads 
\begin{equation}
 \label{equ:self-energy}
{\rm Re}\,\Sigma(p) = -\frac{8T}{\sqrt{\pi}}
    \frac{1}{p}F_D\left(\frac{p}{\sqrt{2mT}}\right) \frac{z}{a} \, ,
\end{equation}
where $F_D(\tilde{p})$ is Dawson's integral and $\tilde{p}=p/\sqrt{2mT}$. 
The momentum-dependence in $\Delta E_p$ modifies the quasi-particle velocity 
as $\vec{v}_p=\vec{\nabla}_p\,E_p=\vec{v}_p^{\,0}+\Delta\vec{v}_p$, where 
$\vec{v}_p^{\,0}=\vec{p}/m$ is the velocity of a free particle and 
\begin{equation}
\label{equ:velocity}
\Delta\vec{v}_p = \frac{\vec{p}}{m} \mathcal{G}(\tilde{p}) \frac{z}{a}, 
 \hspace{1.0cm}
\mathcal{G}(\tilde{p}) = \frac{2}{\pi}\frac{\lambda}{\tilde{p}^3} 
  \left(F_D(\tilde{p})[1+2\tilde{p}^2]-\tilde{p}\right) \, . 
\end{equation}
As a consistency check we can verify that the $(z/a)$-dependence of the 
quasi-particle properties is compatible with the equation of state 
controlled by the second virial coefficient. In kinetic theory the 
enthalpy $\mathcal{E}+P$ can be written as \cite{Schaefer:2013oba}
\begin{equation} 
\label{equ:enthalpy}
\mathcal{E}+P = \nu \int d\Gamma_p \left(
 \frac{1}{3}\vec{p}\cdot\vec{v}_p + E_p \right) f_p \, ,
\end{equation}
where $d\Gamma_p=d^3p/(2\pi)^3$ and $f_p(\vec{x},t)$ is the quasi-particle 
distribution function. From Eq.~(\ref{equ:enthalpy}), we can compute in 
equilibrium the $\mathcal{O}(z(\lambda/a))$-shift in the enthalpy due to the change 
in the quasi-particle energy and velocity discussed above. We get 
\begin{equation} 
\Delta \left( \mathcal{E}+P \right) =
 \frac{2\nu}{3} \int d\Gamma_p\, \frac{p^2}{2m} 
   \left( \frac{\Delta v_p}{v_p^0}\right) \hat{f}_p^{0}
 + \nu \int d\Gamma_p \,\Delta E_p 
  \left( 1 - \frac{5}{3} \frac{p^2}{2mT}\right) \hat{f}_p^{0} \,, 
\end{equation}
where $\hat{f}_p^0=z\,e^{-E_p^0/T}$ is the equilibrium distribution function 
for the non-interacting system. Using Eqs.~(\ref{equ:self-energy}) and
(\ref{equ:velocity}) we find 
\begin{equation} 
\label{equ:enthalpyLO}
\Delta \left( \mathcal{E}+P \right) = \frac{2}{\pi}
    \frac{\lambda}{a} z^2 \frac{\nu T}{\lambda^3} \,. 
\end{equation}
This result can be compared to the virial expansion. We use the thermodynamic 
identity $\mathcal{E}+P=\mu n+ sT$ with $n$ and $s$ given above and determine 
${\cal O}(\lambda/a)$ corrections to the enthalpy from the second virial 
coefficient given in Eq.~(\ref{equ:secondvirialcoefficient}). The result 
agrees with Eq.~(\ref{equ:enthalpyLO}). 

\section{Shear viscosity from kinetic theory \label{sec:3}}
\subsection{Chapman Enskog expansion\label{sec:31}}

 We compute the shear viscosity by matching the expression for the 
dissipative contribution to the stress tensor in fluid dynamics to 
the result in kinetic theory. In fluid dynamics we write 
\begin{equation}
\delta \Pi^{ij} = -\eta \sigma^{ij}-\zeta\delta^{ij} \nabla_ku_k,
\hspace{1cm}
\sigma_{ij} = \nabla_i u_j + \nabla_j u_i - \frac{2}{3} \delta_{ij} 
   \nabla_ku_k\, ,
\end{equation} 
where $\vec{u}$ is the fluid velocity, $\eta$ is the shear viscosity, 
and $\zeta$ is the bulk viscosity. In kinetic theory $\delta\Pi_{ij}$ is expressed 
in terms of the non-equilibrium part $\delta f_p=f_p-f_p^0$ of the distribution
function. We have 
\begin{equation}
\delta \Pi^{ij} = \nu \int d\Gamma_p\, p^i v^j_p \,\delta f_p\, . 
\end{equation}
In the classical limit it is convenient to define an off-equilibrium 
function $\psi_p$ by $f_p=f_p^0(1-\psi_p/T)$ where $f_p^0=z\,e^{-\left(E_p-\vec{p}
\cdot\vec{u}\right)/T}$. The function $\psi_p$ is determined by the 
Boltzmann equation
\begin{equation}
\label{equ:Eq1}
 \mathcal{D}f_p \equiv \left(\frac{\partial}{\partial t} 
        + \vec{v}_p\cdot\vec{\nabla}_x + \vec{F}\cdot\vec{\nabla}_p\right)f_p 
 = \mathcal{C} \,.
\end{equation}
In Eq.~(\ref{equ:Eq1}), $\mathcal{D}f_p$ denotes the streaming term in 
which $\vec{F}=-\vec{\nabla}_x\,E_p$ is the force acting on the quasi-particles
between collisions, and $\mathcal{C}$ is the collision operator term. In 
the high temperature limit the collision term is dominated by two-body 
scatterings. Linearizing in the off-equilibrium function $\psi_p$ we 
get~\cite{Lifshitz} 
\begin{equation}
 \label{equ:Eq2}
 \mathcal{C} = \frac{f^0_{p_1}}{T} \int \prod_{i=2}^4 d\Gamma_{p_i}\,
  f^0_{p_2} w(1,2;3,4) \left( \psi_{p_1}+\psi_{p_2}-\psi_{p_3}-\psi_{p_4}\right)\, ,
\end{equation}
where $w(1,2;3,4)=(2\pi)^4\,\delta^{(3)}(\vec{p_1}+\vec{p_2}-\vec{p_3}-\vec{p_4})
\,\delta(E_{p_1}+E_{p_2}-E_{p_3}-E_{p_4})|\mathcal{A}|^2$ is the transition rate
and $|\mathcal{A}|^2$ is the absolute square of the scattering amplitude. 

 In order to determine the shear viscosity we expand $\psi_p$ in gradients 
of the thermodynamic variables. This is known as the Chapman-Enskog expansion.
At linear order in gradients of $\vec{u}$ we can write $\psi_p = \chi^{ij}(p)\,
\sigma_{ij}$, where we have dropped terms that contribute to the bulk viscosity 
\cite{Schaefer:2013oba}. In the local rest-frame of the fluid we find
\begin{equation}
 \label{equ:Eq3}
 \frac{T}{f^0_{p_1}}(\mathcal{D} f^0_{p_1}) 
   =  \frac12 v_{p_1}^i p_1^j\,\sigma_{ij} 
   = \frac{1}{2m} p_1^i p_1^j \left(1+
             \mathcal{G}(\tilde{p}_1)\frac{z}{a}\right)\sigma_{ij} 
   = \mathcal{C}[\chi^{ij}(p_1)]\,\sigma_{ij} \,,
\end{equation}
where, using Eq.~(\ref{equ:Eq2}), 
\begin{equation}
 \label{equ:Eq4}
 \mathcal{C}[\chi^{ij}(p_1)]\,\sigma_{ij} 
   = \int \prod_{i=2}^4 d\Gamma_{p_i} f^0_{p_2} w(1,2;3,4) 
           \left(\chi^{ij}(p_1)+\chi^{ij}(p_2)
                 -\chi^{ij}(p_3)-\chi^{ij}(p_4)\right)\sigma_{ij} \,.
\end{equation}
The linearized Boltzmann equation in the shear channel can be written as
\begin{equation}
 \label{equ:Eq6}
 \frac{1}{2m} \Big\langle\chi^{ij}(p_1)\Big|
     (p_1)_{ij}\left(1+\mathcal{G}(\tilde{p}_1)\frac{z}{a}\right)\Big\rangle 
  = \Big\langle\chi^{ij}(p_1)\Big|
     \mathcal{C}\left[\chi_{ij}(p_1)\right]\Big\rangle \,,
\end{equation}
where $p_{ij}=p_ip_j-\frac13\delta_{ij}\,p^2$ with $p^{ij}\sigma_{ij}=p^ip^j
\sigma_{ij}$. We have also defined an inner product on the space of 
linearized distribution functions
\begin{equation}
\langle a(p)|b(p)\rangle=\int d\Gamma_p\, f_p^0\,a(p) b(p)\, . 
\end{equation}
The shear part of the stress tensor can be written as 
\begin{equation}
 \label{equ:Eq8}
 \delta \Pi^{ij} = -\frac{\nu}{mT} \int d\Gamma_{p_1} 
     \,f_{p_1}^0 \chi(p_1) \left(1+\mathcal{G}(\tilde{p}_1)\frac{z}{a}\right) 
    p_1^i p_1^j p_1^{kl} \sigma_{kl} \, ,
\end{equation}
where we have defined $\chi^{ij}(p)=p^{ij}\chi(p)$. Using $p^{ij}p_{ij} = p^ip^j
p_{ij} = \frac23 p^4$ and Eq.~(\ref{equ:Eq6}) we finally obtain
\begin{equation}
 \label{equ:Eq9}
 \eta = \frac{\nu}{10\,m^2T}\frac{\Big\langle\chi^{ij}(p_1)\Big|(p_1)_{ij}
         \left(1+\mathcal{G}(\tilde{p}_1)\frac{z}{a}\right)
          \Big\rangle^2}{\Big\langle\chi^{ij}(p_1)\Big|
               \mathcal{C}\left[\chi_{ij}(p_1)\right]\Big\rangle}\,.
\end{equation}
This expression determines $\eta$ for a given off-equilibrium function
$\chi(p)$ that solves the linearized Boltzmann equation. By writing the
result in the specific form given in Eq.~(\ref{equ:Eq9}) we also obtain
a lower bound on the shear viscosity for any trial function $\chi(p)$. 
The result for $\eta$ can then be found by maximizing Eq.~(\ref{equ:Eq9})
over all trial functions. For simplicity, we will use $\chi(p)=1$ as a
trial function. This function was shown to provide an excellent approximation,
accurate to better than 2\%, for the shear viscosity at 
unitarity~\cite{Bruun:2006kj}. 

\subsection{In-medium cross section\label{sec:32}}

 Medium effects influence the shear viscosity in a variety of ways. The
medium modification of the quasi-particle velocity impacts the streaming term 
in Eq.~(\ref{equ:Eq3}), and the stress tensor in Eq.~(\ref{equ:Eq8}).
Both of these contribute to the numerator in Eq.~(\ref{equ:Eq9}). The matrix 
element of the collision operator in the denominator is affected by medium 
modifications of the quasi-particle energy that enters into the distribution 
functions and the transition rate $w(1,2;3,4)$. Moreover, medium effects 
modify also the squared scattering amplitude $|\mathcal{A}|^2$.

 In order to explore these effects we begin with a simple model calculation
in which we take into account medium effects in the scattering amplitude, and thus in the cross section, only. 
The absolute square of the vacuum scattering amplitude is
\begin{equation}
 \label{equ:14}
 |\mathcal{A}|^2 = \frac{16\pi^2}{m^2}\frac{a^2}{a^2q^2+1} \, ,
\end{equation}
where $\vec{q}=(\vec{p}_2-\vec{p}_1)/2$ is the relative momentum between two scattering
particles. In terms of the zero range Lagrangian given in Eq.~(\ref{L_4f}) 
the amplitude arises from the sum of all two-body scattering diagrams. 
These diagrams form a geometric series and $\mathcal{A}=C_0/(1-\Pi_0 C_0)$,
where $\Pi_0$ is the two-particle polarization function. In dimensional 
regularization we find $\Pi_0(q)=-imq/(4\pi)$ and $C_0=4\pi a/m$. At leading order
in the fugacity medium effects arise from Pauli-blocking of the fermion
lines in $\Pi_0$. We can write $\Pi=\Pi_0+\delta\Pi$ 
with~\cite{Bruun:2005en,Bruun:2006kj}
\begin{equation}
 \label{equ:15}
 \delta\Pi(P,q) = -\int \frac{d^3k}{(2\pi)^3} 
     \frac{\hat{f}^0_{|\vec{P}/2+\vec{k}|}+\hat{f}^0_{|\vec{P}/2-\vec{k}|}}
          {(q^2-k^2)/m+i\epsilon}\, ,
\end{equation}
which depends on both the relative momentum $q$ and the total momentum 
$\vec{P}=\vec{p}_1+\vec{p}_2$. The  real and imaginary parts of the 
in-medium polarization function are given by  
\begin{eqnarray}
 \label{equ:ImDeltaPi}
 {\rm Im}\,\delta\Pi & = & \frac{z}{\pi} m^2 T 
         \frac{e^{-P^2/(8mT)}}{P} e^{-q^2/(2mT)} \sinh(Pq/(2mT)) \,, \\
 \label{equ:ReDeltaPi}
 {\rm Re}\,\delta\Pi & = & -\frac{2z}{\pi^2} m^2 T 
         \frac{e^{-P^2/(8mT)}}{P} \int_0^\infty dx \frac{x e^{-x^2} 
            \sinh(Px/\sqrt{2mT})}{(q^2/(2mT)-x^2)} \,,
\end{eqnarray}
where the integral in Eq.~(\ref{equ:ReDeltaPi}) 
is a Cauchy principle value 
integral. The full in-medium scattering amplitude squared is 
\begin{equation}
 \label{equ:16}
|\mathcal{A}|^2 = \frac{16\pi^2}{m^2}
  \frac{1}{\left(q-\frac{4\pi}{m}\,{\rm Im}\,\delta\Pi\right)^2
          +\left(\frac{1}{a}-\frac{4\pi}{m}\,{\rm Re}\,\delta\Pi\right)^2}  \, ,
\end{equation}
which agrees with the ``broad-resonance'' expression discussed 
in~\cite{Bruun:2005en}. Phenomenological consequences of this result
were also discussed in~\cite{Bruun:2005en}. The important observation in 
our context is that, whereas the squared vacuum amplitude is even in $a$, the 
in-medium expression has odd corrections of order $\mathcal{O}(z(a/\lambda))$.

 The calculation of the shear viscosity based on the in-medium scattering
amplitude is now straightforward. We define the total and relative momenta 
in the initial and final state as $\vec{P}=\vec{p}_1+\vec{p}_2$ and 
$\vec{P}'=\vec{p}_3+\vec{p}_4$, as well as $\vec{q}=(\vec{p}_1-\vec{p}_2)/2$ 
and $\vec{q}\,'=(\vec{p}_3-\vec{p}_4)/2$. The vector $\vec{P}$ can be 
aligned along the $z$-axis, and the integration over $d^3P'$ is performed 
by using the condition for total momentum conservation. This leaves 
three angular integrals, $d\cos\theta_q\, d\cos\theta_{q'}\, d\phi$, 
where 
\begin{equation}
\vec{P}\cdot\vec{q} = Pq\cos\theta_q, \hspace{0.25cm}
\vec{P}\cdot\vec{q}\,' = Pq'\cos\theta_{q'}, \hspace{0.25cm}
\vec{q}\cdot\vec{q}\,' = qq'\left[ \cos\theta_q\cos\theta_{q'}
                +\sin\theta_q\sin\theta_{q'}\cos\phi\right]. 
\end{equation}
Finally, the integration over $dq'$ can be performed by making use of 
the condition for energy conservation $q^2/m-q'^2/m=0$. Inside the integral
the off-equilibrium factor $\chi^{ij}(p_1)\left(\chi^{ij}(p_1)+\chi^{ij}(p_2)
-\chi^{ij}(p_3)-\chi^{ij}(p_4)\right)$ can be symmetrized in the in- and
out-going momenta. We find
\begin{equation}
\frac14 \left(\chi^{ij}(p_1)+\chi^{ij}(p_2)
          -\chi^{ij}(p_3)-\chi^{ij}(p_4)\right)^2
 =  q^4+q'^4-\frac13(q^2-q'^2)^2-2q^2q'^2\cos^2\Theta\, , 
\end{equation}
where $\vec{q}\cdot\vec{q}\,'=qq'\cos\Theta$. The matrix element of
the linearized collision operator is then given by 
\begin{equation}
\label{equ:CollMatrixElement}
 \langle\chi^{ij}|\mathcal{C}[\chi_{ij}]\rangle 
    = z^2 \frac{m}{6\pi^5} \int_0^\infty dP \int_0^\infty dq \, 
       P^2 q^7 e^{-P^2/(4mT)} e^{-q^2/(mT)} |\mathcal{A}|^2 \,.
\end{equation}
The remaining integrations can be performed numerically. The integral in 
the numerator of Eq.~(\ref{equ:Eq9}) is $\langle\chi^{ij}(p)|p_{ij}\rangle
=5 z (mT)^{7/2}/\sqrt{2\pi^3}$.

\begin{figure}[t]
\centering
\includegraphics[scale=0.32]{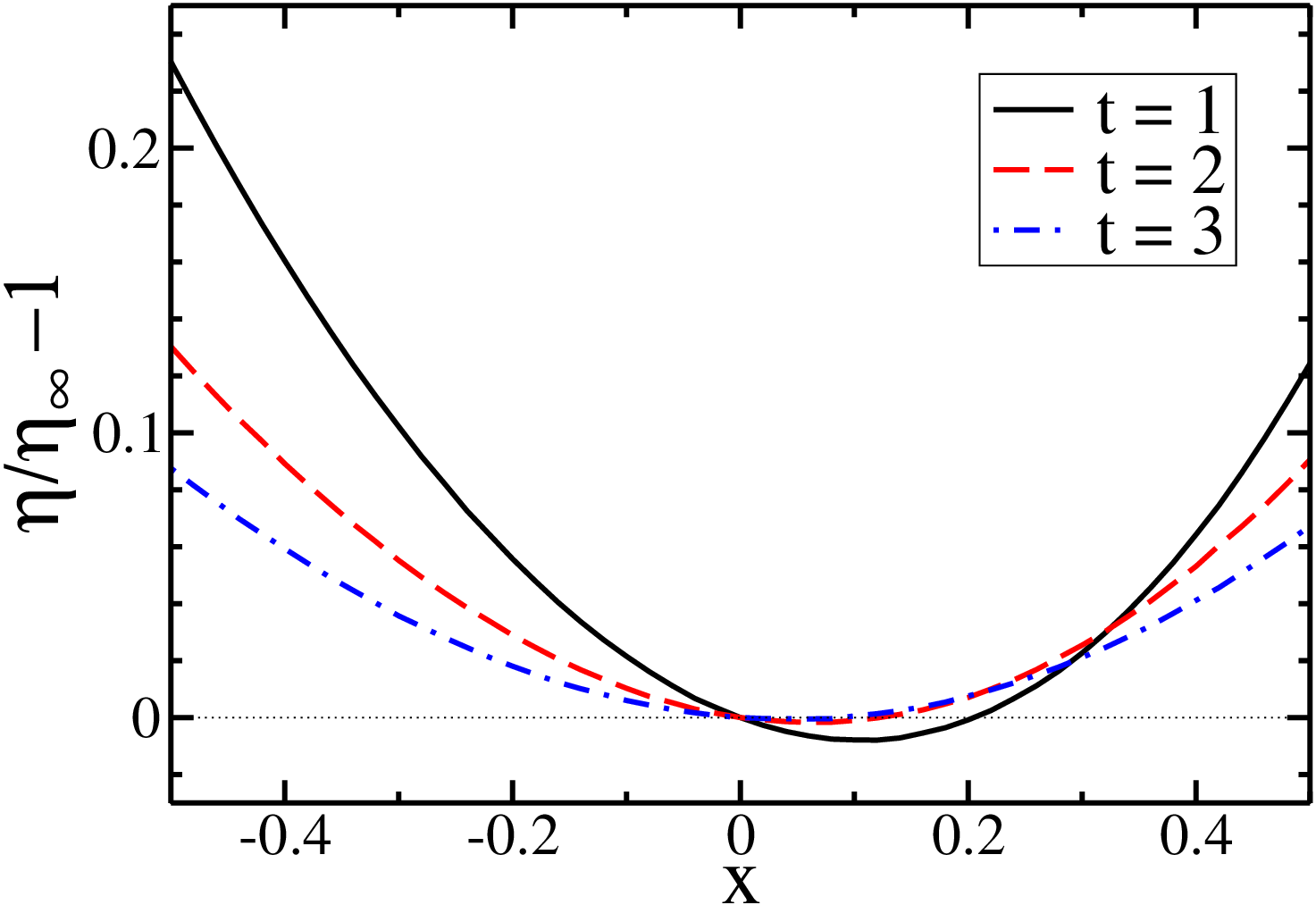} 
\hspace{2mm}
\includegraphics[scale=0.32]{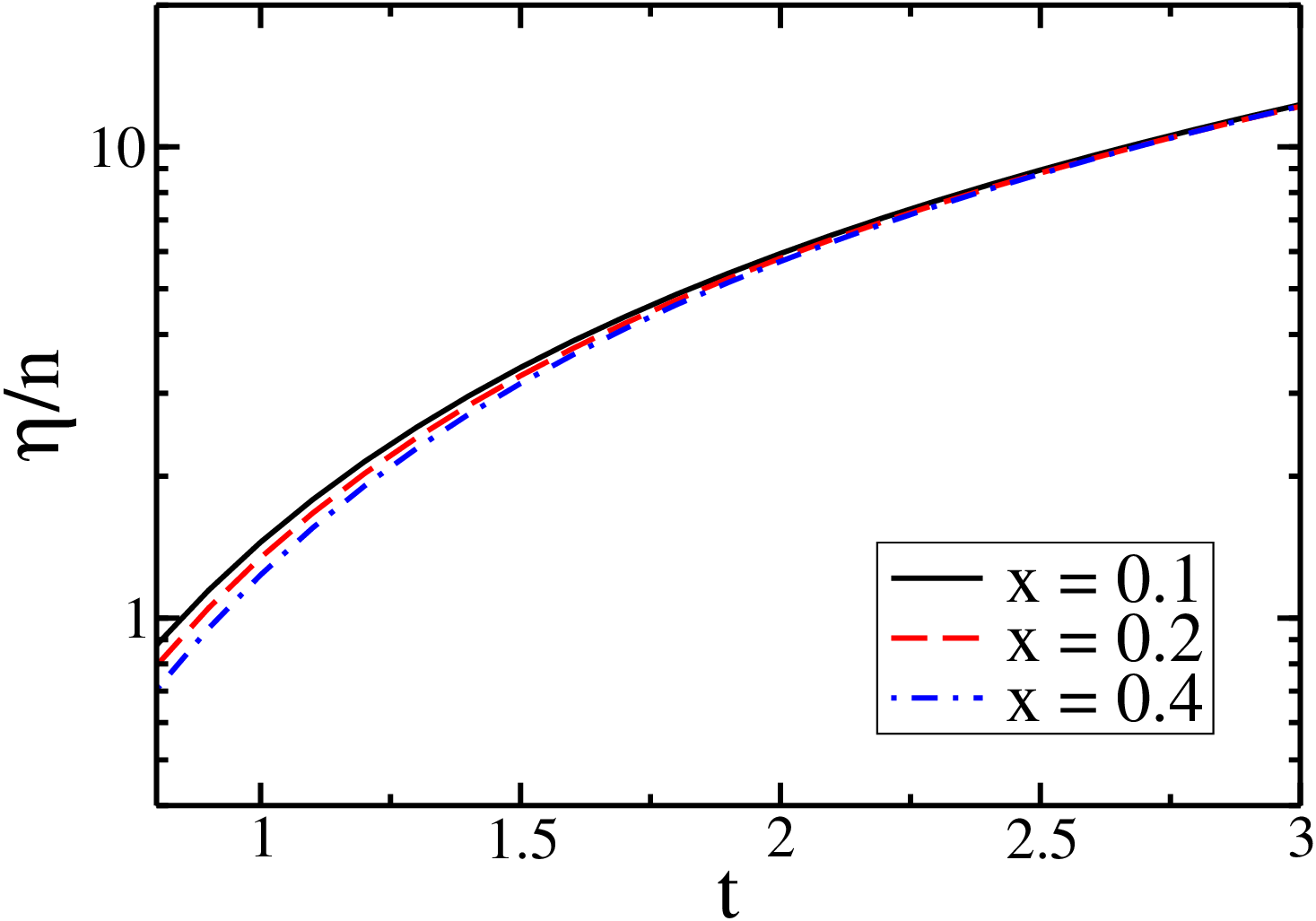} 
\caption[]{\label{fig:fig1} 
(Color online) Left panel: Scaled shear viscosity difference from unitarity 
$(\eta-\eta_\infty)/\eta_\infty$ as a function of $x=1/(k_Fa)$ for different 
values of $t=T/T_F$. The shear viscosity is computed from the 
in-medium cross section containing only the influence of ${\rm Re}\,\delta\Pi$. 
Furthermore, medium corrections to the quasi-particle
energy and velocity are neglected.
Right panel: Shear viscosity $\eta$ scaled by the particle density $n$ 
from Eq.~(\ref{equ:particledensity}) as a function of $t$ for different 
values of $x$.}
\end{figure}

 Numerical results are shown in Fig.~\ref{fig:fig1}. In the left panel 
we focus on the difference $\eta-\eta_\infty$, where $\eta_\infty$ is the 
shear viscosity at unitarity. In this difference medium corrections that 
are independent of $a$ cancel. We plot the dimensionless quantity $\eta/
\eta_\infty -1$ as a function of $1/(k_Fa)$ for different values of $T/T_F$. 
In the right panel of Fig.~\ref{fig:fig1} we show the behavior of $\eta/n$ 
as a function of $T/T_F$ for three different values of $1/(k_Fa)$
on the BEC side.

 Our results can be compared to recent measurements reported by the 
North Carolina State University group~\cite{Elliott:2014nn}. Elliott et al.~studied
dissipative corrections to the expansion of a dilute Fermi gas 
for several different values of $1/(k_Fa)$ near unitarity. The results are 
reported in terms of a trap averaged kinematic shear viscosity $\langle
\eta/n\rangle$. It is difficult to compare these results directly to 
our calculations for a homogeneous gas because performing the trap 
average involves a poorly constrained cutoff on the spatial integral 
over the shear viscosity. However, the results of Elliott 
et al.~are quite remarkable, even on a qualitative level. 
They find that $\langle\eta/n\rangle$
has a minimum on the BEC-side of the resonance. As the temperature of the
cloud increases, this minimum is shifted toward the unitarity limit.
At a given value of $k_Fa$, $\langle\eta/n\rangle$ increases
with temperature on the BEC side, and decreases on the BCS side.
This behavior is consistent with a $(\lambda/a)^2$ dependence that dominates
at high temperature, and competes with a $z(\lambda/a)$ contribution
that becomes more important as the temperature is lowered.
On a more quantitative level, Elliott et al.~study the expanding
gas at a cloud energy $\tilde{E}/E_F=1$ and find a minimum of 
$\langle\eta/n\rangle$ at $1/(k_Fa)\simeq 0.18$. Here, $\tilde{E}$ 
is a virial theorem based measure of the cloud energy. For exactly 
harmonic traps $\tilde E=E_{tot}/N$, where $E_{tot}$ is the total energy
(internal plus potential) of the cloud. In the high temperature limit
we expect $E_{tot}=3NT$.

 The results shown in Fig.~\ref{fig:fig1} are consistent with 
these findings. 
The figure in the left panel demonstrates that we observe a minimum of 
$\eta/\eta_\infty-1$ on the BEC side. The minimum shifts toward unitarity 
with increasing temperature. Moreover, on the BCS side the scaled 
difference decreases with increasing temperature, while on the BEC side, 
close to unitarity, it increases with increasing $T$. The figure in the
right panel shows that $\eta/n$ is independent of $1/(k_Fa)$ for large 
$T/T_F$, and that the sensitivity to $1/(k_Fa)$ grows with decreasing 
temperature. On the BEC side $\eta/n$ drops with increasing $1/(k_Fa)$ even
beyond the point where a minimum was observed in the left panel. This 
is a consequence of the decrease of $n$ with increasing 
$1/(k_Fa)$ as predicted by Eq.~(\ref{equ:particledensity}).

\subsection{Expansion in $z(\lambda/a)$ \label{sec:33}}

As discussed above, in-medium effects influence the shear viscosity in 
several ways. In addition to the effects of the in-medium scattering 
amplitude, the medium modification of the quasi-particle energy affects the 
energy conserving delta function and the final state momenta. With
the interaction included, energy conservation implies 
$q^2-q'^2=m\mathcal{F}(P,q^2,q'^2,\theta_q,\theta_{q'})$ with
\begin{eqnarray}
\mathcal{F}&=&\Delta E_p(P,q'^2,\cos\theta_{q'})
           +\Delta E_p(P,q'^2,-\cos\theta_{q'}) \nonumber \\
 && \mbox{}           
           -\Delta E_p(P,q^2,\cos\theta_q)
           -\Delta E_p(P,q^2,-\cos\theta_q)\, ,
\end{eqnarray}
see Appendix~\ref{App:A} for details. We can solve this equation for 
$q'^2$ order-by-order in the fugacity and perform the integration over $dq'$ 
in the matrix element of the collision operator up to order $\mathcal{O}(z)$. This amounts to
\begin{eqnarray}
 \nonumber
 \langle\chi^{ij}|\mathcal{C}[\chi_{ij}]\rangle 
  & = & \frac{2}{(2\pi)^6} 
         \int_{0}^{\infty} dP \int_{0}^{\infty} dq \int_{-1}^{1} d\cos\theta_q 
         \int_{-1}^{1} d\cos\theta_{q'} \int_{0}^{2\pi} d\phi \;
             P^2 q^2 \,  f_{p_1}^0 f_{p_2}^0 \, |\mathcal{A}|^2 \\ 
 \label{equ:fullColl}
 & & \hspace{-5mm} 
   \times \left\{mq^5(1-\cos^2\!\Theta)
 \left(1-\left.m\frac{\partial\mathcal{F}}{\partial q'^2}
                                        \right|_{q'^2=q^2}\right)
    +\frac32 mq^3(1-\cos^2\!\Theta)\,\Delta(q'^2)\right\} \,,
\end{eqnarray}
where $f_p^0=z\,e^{-E_p/T}$ contains medium effects through $\Delta E_p$ 
and $\Delta(q'^2)=-\left.m\mathcal{F}\right|_{q'^2=q^2}$. Medium corrections 
to the numerator of Eq.~(\ref{equ:Eq9}) arise from modifications of the 
quasi-particle energy and velocity. We find
\begin{equation}
 \label{equ:fullNumerator}
   \langle\chi^{ij}(p)|p_{ij}\left(1+\mathcal{G}(\tilde{p})z/a\right)\rangle
 = \frac{z (2mT)^{7/2} }{3\pi^2}
      \int_0^\infty dy \, y^6 e^{-y^2} e^{-{\rm Re}\,\Sigma(\sqrt{2mT}y)/T} 
          \left\{1+\mathcal{G}(y)\frac{z}{a}\right\} \,.
\end{equation}

\begin{figure}[t]
\centering
\includegraphics[scale=0.32]{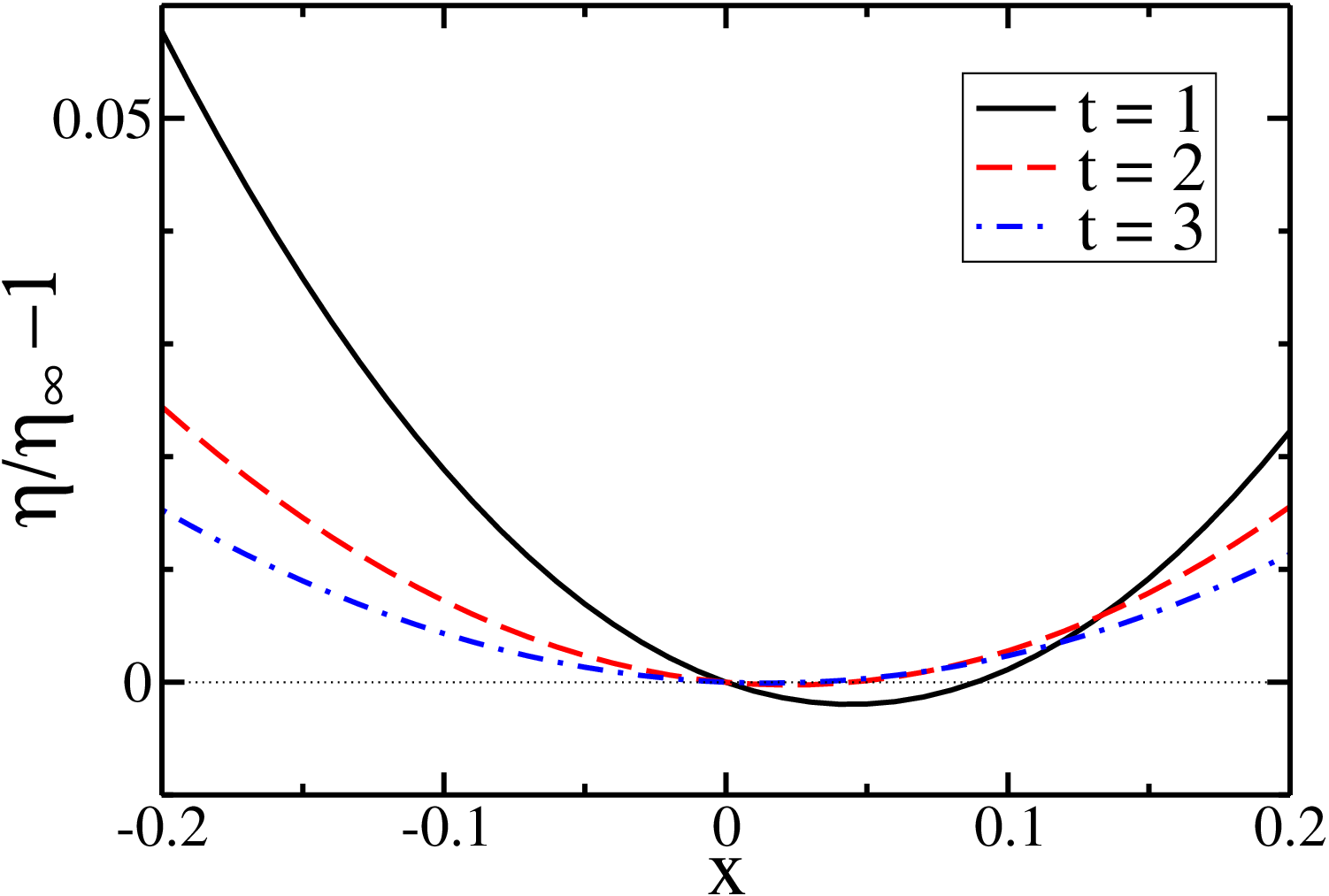} 
\caption[]{\label{fig:fig2} 
(Color online) Scaled shear viscosity difference $(\eta-\eta_\infty)/\eta_\infty$ 
as a function of $x=1/(k_Fa)$ for different values of $t=T/T_F$. This 
figure shows the result of a systematic expansion to order $\mathcal{O}((\lambda/a)^2)$
and $\mathcal{O}(z(\lambda/a))$, see Eq.~(\ref{equ:Expansion}).}
\end{figure}

Equations~(\ref{equ:fullColl}) and~(\ref{equ:fullNumerator}) contain all 
the terms needed to compute the dependence of $\eta$ on $(\lambda/a)$ 
at leading order in $z$. Close to unitarity we can expand 
\begin{equation}
 \label{equ:Expansion}
 \frac{\eta-\eta_\infty}{\eta_\infty} 
       = c_0 \left(\frac{\lambda}{a}\right)^2 
         + c_1\, z \left(\frac{\lambda}{a}\right) + \dots \, . 
\end{equation}
Details of the calculation of $c_0$ and $c_1$ are described in 
Appendix~\ref{App:B}. We find 
\begin{equation}
c_0 =  \frac{1}{4\pi}\simeq 0.07958\, , \hspace{1cm}
c_1 \simeq  -0.03325 \, , 
\end{equation}
where the value of $c_1$ is the result of a numerical calculation. The final 
result for $c_1$ involves subtle cancellations between several effects. We
showed in the previous section that the in-medium cross section 
alone leads to a minimum of the shear viscosity 
on the BEC side, corresponding to a negative contribution to 
$c_1$. In contrast, corrections to the quasi-particle 
velocity give a positive contribution to $c_1$. 
This effect is largely cancelled by corrections to the quasi-particle energy, 
see Appendix~\ref{App:B}. The final result 
of Eq.~(\ref{equ:Expansion}) is shown in Fig.~\ref{fig:fig2}. We observe 
that the complete result to ${\cal O}((\lambda/a)^2)$ and 
${\cal O}(z(\lambda/a))$ is remarkably similar to Fig.~\ref{fig:fig1}, 
which only includes the in-medium cross section.

\section{Conclusions and outlook \label{sec:4}}

 In summary, we studied the influence of in-medium effects on the 
scattering length dependence of the shear viscosity in the dilute Fermi 
gas near unitarity. To zeroth order in the fugacity, $\eta$ only 
depends on $(\lambda/a)^2$, and the minimum occurs at unitarity. Medium 
effects give, however, corrections of order $\mathcal{O}(z(\lambda/a))$, and the minimum 
of $(\eta-\eta_\infty)/\eta_\infty$ shifts to the BEC side. The main 
effect that causes this behavior is Pauli blocking in the in-medium 
cross section. Our results are in qualitative agreement with the 
experimental observations recently reported in~\cite{Elliott:2014nn}. 
More detailed comparisons will require an improved understanding 
of how to average the shear viscosity over the trap. 

 In our calculation we focused on one- and two-body effects, and 
truncated the systematic expansion in $z$ and $(\lambda/a)$ at 
order $\mathcal{O}((\lambda/a)^2)$ and $\mathcal{O}(z(\lambda/a))$. As demonstrated in 
Sect.~\ref{sec:32}, a simple model involving only in-medium scattering
can be studied at any value of $z$ and $(\lambda/a)$, although the 
applicability of kinetic theory becomes questionable for $z\gtrsim 1$
or $(\lambda/a)\gtrsim 1$. Note that the former condition arises from
the applicability of kinetic theory in a homogeneous system, 
whereas the latter condition arises in the context of applying 
kinetic theory to a finite system, in which the mean free path
must be small compared to the system size. We have not considered 
three-body collisions, which are expected to contribute to the shear 
viscosity at ${\cal O}(z)$. Vacuum terms in the three-body amplitude 
can contain terms of ${\cal O}(1/(qa))$, which would contribute to 
the shear viscosity at ${\cal O}(z(\lambda/a))$. There is no 
experimental evidence for three-body effects in transport coefficients, 
but a theoretical estimate is certainly desirable. 

 The results presented in this work rely on kinetic theory and 
are limited to temperatures significantly above the phase 
transition. Near $T_c$ quantum statistics and pseudogap effects
are likely to be important. These effects can be incorporated
into kinetic theory, but a diagrammatic framework along the 
lines of \cite{Enss:2010qh,Guo:2010,He:2014xsa} is likely to 
be more reliable. 

\section*{Acknowledgements}

 We would like to thank J.~A.~Joseph and J.~E.~Thomas for many
fruitful discussions regarding their experiments. We also acknowledge valuable 
comments by G.~M.~Bruun regarding our work. 
This work is supported by the US Department of Energy 
grant DE-FG02-03ER41260. 

\appendix
\numberwithin{equation}{section}
\section{Condition for energy conservation \label{App:A}}

The matrix element of the collision operator in Eq.~(\ref{equ:Eq9}) can 
be written as 
\begin{eqnarray}
\nonumber
 \langle\chi^{ij}|\mathcal{C}[\chi_{ij}]\rangle 
   & = & \frac{2}{(2\pi)^6} 
          \int_0^\infty dP \int_0^\infty dq 
          \int_{-1}^{1} d\cos\theta_q \int_{-1}^{1} d\cos\theta_{q'} \int_0^{2\pi} d\phi 
          \,P^2 q^2 \, f_{p_1}^0 f_{p_2}^0 \, 
         |\mathcal{A}|^2 \\ 
 \label{equ:A1}
 & & \hspace{-19mm} 
 \times \int_0^\infty dq' q'^2 
      \left[q^4+q'^4-2q^2q'^2\cos^2\!\Theta-\frac13(q^2-q'^2)^2\right]
       \delta\left(q^2/m-q'^2/m-\mathcal{F}\right) \,.
\end{eqnarray}
We change variables from $dq'$ to $dq'^2$ and use the energy conserving 
delta function to evaluate the $q'^2$-integral up to leading order in $z$. 
For this purpose we solve the condition $q^2/m-q'^2/m=\mathcal{F}(P,q^2,
q'^2,\theta_q,\theta_{q'})$ for $q'^2$ order-by-order in the fugacity. 
Since $m\mathcal{F}$ is of order $\mathcal{O}(z(\lambda/a))$, we can replace 
$q'^2$ in $\mathcal{F}$ by the solution at $\mathcal{O}(z^0)$, i.e.~$q'^2=q^2$, 
and find $q'^2=q^2\left(1+\Delta(q'^2)/q^2\right)$ 
with the $z(\lambda/a)$-correction 
\begin{eqnarray}
\nonumber
 \Delta(q'^2) & = & 
     -\left.m\mathcal{F}\right|_{q'^2=q^2} 
      \,=\, m \bigg(\Delta E_p(P,q^2,\cos\theta_q)
                   +\Delta E_p(P,q^2,-\cos\theta_q) \\
 \label{equ:A2}
 & & \hspace{5mm} 
                   -\Delta E_p(P,q^2,\cos\theta_{q'})
                   -\Delta E_p(P,q^2,-\cos\theta_{q'})\bigg) \,,
\end{eqnarray}
where 
\begin{eqnarray}
 \label{equ:A3}
 \Delta E_p(\alpha,\beta,\gamma) & = & 
       -\frac{8T}{\sqrt{\pi}} \frac{1}{p(\alpha,\beta,\gamma)} 
  F_D\left(\frac{p(\alpha,\beta,\gamma)}{\sqrt{2mT}}\right) \frac{z}{a} \,, \\
 \label{equ:A4}
 p(\alpha,\beta,\gamma) & = & \sqrt{\frac{\alpha^2}{4}
                  +\beta+\alpha\,\gamma\sqrt{\beta}}
\end{eqnarray}
such that $\Delta(q'^2)=0$ for $q^2=0$. With this, the integral over 
$dq'^2$ can be evaluated. We find 
\begin{eqnarray}
 \nonumber
 & & 
 \frac{1}{2} \int_0^\infty dq'^2 \sqrt{q'^2} 
      \left[q^4+q'^4-2q^2q'^2\cos^2\!\Theta-\frac13(q^2-q'^2)^2\right]
      \delta\left(q^2/m-q'^2/m-\mathcal{F}\right) \\
 \label{equ:A5}
 & & \hspace{-8mm}
   = mq^5(1-\cos^2\!\Theta)-m^2q^5(1-\cos^2\!\Theta)
     \left.\frac{\partial\mathcal{F}}{\partial q'^2}\right|_{q'^2=q^2} 
  + \frac{3}{2} mq^3(1-\cos^2\!\Theta)\Delta(q'^2)+\dots \,,
\end{eqnarray}
where the first term is of $\mathcal{O}(z^0)$ and the second and third term 
are of $\mathcal{O}(z)$. Since $\partial\mathcal{F}/\partial q'^2$ 
is already of $\mathcal{O}(z)$, it suffices to evaluate this factor 
at $q'^2=q^2$. 

\section{Systematic expansion \label{App:B}}

 Ignoring all medium effects in Eq.~(\ref{equ:Eq9}), in particular 
using Eq.~(\ref{equ:14}) for the squared scattering amplitude, one obtains 
for the shear viscosity 
\begin{equation}
 \label{equ:B1}
 \eta_0 = \frac{15}{2^8\sqrt{\pi}}(mT)^{3/2} \mathcal{J}^{-1}
\end{equation}
with 
\begin{equation}
 \label{equ:B2}
 \mathcal{J} = \int_0^\infty d\tilde{q} \,
               \frac{\tilde{q}^5e^{-2\tilde{q}^2}}{1+1/(2mTa^2\tilde{q}^2)} 
  = \frac18 \left(1-\frac{1}{2mTa^2}+\dots\right) \,,
\end{equation}
where $\tilde{q}=q/\sqrt{2mT}$. In-medium corrections alter this result 
as $\eta=\eta_0+\Delta\eta$. Since $\eta$ in Eq.~(\ref{equ:Eq9}) is of 
the form $\eta=A/B$, one can determine $\Delta\eta$ to leading order 
in the deviations from $\eta_0=A_0/B_0$ as $\Delta\eta=\Delta A/B_0
-\eta_0\Delta B/B_0$. At unitarity, one finds from Eq.~(\ref{equ:Eq9}) 
to leading order in $z$ 
\begin{equation}
 \label{equ:EtaInfty}
 \eta_\infty = \frac{15\pi}{8\sqrt{2}}\frac{1}{\lambda^3}
    \left(1-\frac{2^5\sqrt{2}}{\sqrt{\pi}}\, z 
   \int_0^\infty d\tilde{P} \int_0^\infty d\tilde{q}\, 
       \tilde{P}\tilde{q}^4\, 
       e^{-3\tilde{P}^2/4}e^{-3\tilde{q}^2}\sinh(\tilde{P}\tilde{q})\right) \, .
\end{equation}
The second term is associated with ${\rm Im}\,\delta\Pi$ in $|\mathcal{A}|^2$ 
and leads to in-medium corrections that cancel in the difference $\eta-
\eta_\infty$. 

 The term $\Delta A$ arises from the change in the quasi-particle velocity 
as well as the energy which enters the distribution function $f_p^0$ inside 
the numerator-integral. We find to leading order 
\begin{equation}
 \label{equ:B3}
 \frac{(\Delta A/B_0)}{\eta_\infty} 
   = \frac{8}{3\pi}\, z\left(\frac{\lambda}{a}\right) \,,
\end{equation}
which gives a non-vanishing contribution to the scaled difference $(\eta-
\eta_\infty)/\eta_\infty$. This tends to decrease $(\eta-\eta_\infty)/\eta_\infty$ 
for $a<0$ and to increase it for $a>0$, i.e.~to shift the minimum to the 
atomic side of the resonance. 

 The term $\Delta B$ is associated with the collision operator. The 
real part of $\delta\Pi$ gives a contribution of $\mathcal{O}(z(\lambda/a))$ 
which is positive for $a<0$ and negative for $a>0$. The imaginary part
of $\delta\Pi$ contributes at $\mathcal{O}(z(\lambda/a)^2)$ in the scaled difference, which we 
have neglected throughout. The distribution functions $f_{p_1}^0 f_{p_2}^0$ in 
Eq.~(\ref{equ:fullColl}) give an order $\mathcal{O}(z(\lambda/a))$ term 
which is positive for $a<0$ and negative for 
$a>0$. Finally, the medium corrections to the energy conserving delta
function give rise to two terms, cf.~Eq.~(\ref{equ:A5}), which are 
both of $\mathcal{O}(z(\lambda/a))$ and decrease (increase) the scaled 
difference for $a<0$ ($a>0$). 

With this, the scaled difference can be expanded systematically in powers 
of $z$ and $(\lambda/a)$ as 
\begin{equation}
 \label{equ:B4}
 \frac{\eta-\eta_\infty}{\eta_\infty} 
   =   c_0 \left(\frac{\lambda}{a}\right)^2 
     + c_1\, z \left(\frac{\lambda}{a}\right) + \dots 
\end{equation}
with $c_0=1/(4\pi)$ and 
\begin{equation}
 \label{equ:B5}
 c_1 = \frac{8}{3\pi}
      +\frac{32\sqrt{2}}{\pi^2}\mathcal{I}_1
      -\frac{16\sqrt{2}}{\pi^{3/2}}\mathcal{I}_2
      -\frac{9}{\sqrt{2}\pi^{5/2}}\mathcal{I}_3
      -\frac{4\sqrt{2}}{\pi^{3/2}}\mathcal{I}_4 \,.
\end{equation}
In Eq.~(\ref{equ:B5}), the integrals are given by 
\begin{eqnarray}
 \label{equ:B6}
 \mathcal{I}_1 & = & 
    \int_0^\infty d\tilde{P} \int_0^\infty d\tilde{q}\, 
           \tilde{P}\tilde{q}^3\, e^{-3\tilde{P}^2/4}e^{-2\tilde{q}^2} 
    \int_0^\infty dx\, \frac{xe^{-x^2}\sinh(\tilde{P}x)}{(\tilde{q}^2-x^2)} \,, \\
 \nonumber
 \mathcal{I}_2 & = & 
    \int_0^\infty d\tilde{P} \int_0^\infty d\tilde{q} \int_0^\pi d\theta_q\, 
     \tilde{P}^2\tilde{q}^5\sin\theta_q\, e^{-\tilde{P}^2/2}e^{-2\tilde{q}^2} \\ 
 \label{equ:B7}
 & & \hspace{2cm} 
    \times \left[\frac{F_D(p(\tilde{P},\tilde{q}^2,\cos\theta_q))}
                      {p(\tilde{P},\tilde{q}^2,\cos\theta_q)}
                +\frac{F_D(p(\tilde{P},\tilde{q}^2,-\cos\theta_q))}
                      {p(\tilde{P},\tilde{q}^2,-\cos\theta_q)}\right] \,, \\
 \nonumber
 \mathcal{I}_3 & = & 
    \int_0^\infty d\tilde{P} \int_0^\infty d\tilde{q} \int_0^\pi d\theta_q 
    \int_0^\pi d\theta_{q'} \int_0^{2\pi} d\phi\, 
    \tilde{P}^2\tilde{q}^3\sin\theta_q\sin\theta_{q'}\,(1-\cos^2\!\Theta) \\
 \label{equ:B8} 
 & & \hspace{2cm} 
        \times\, e^{-\tilde{P}^2/2}e^{-2\tilde{q}^2} 
         \Delta C(\tilde{P},\tilde{q},\theta_q,\theta_{q'}) \,, \\
 \label{equ:B9}
 \mathcal{I}_4 & = & 
         \int_0^\infty d\tilde{P} \int_0^\infty d\tilde{q} 
         \int_0^\pi d\theta_{q'}\, \tilde{P}^2\tilde{q}^4
            \sin\theta_{q'}\, e^{-\tilde{P}^2/2}e^{-2\tilde{q}^2} 
           C'(\tilde{P},\tilde{q},\theta_{q'})
\end{eqnarray}
with $p(\alpha,\beta,\gamma)$ defined in Eq.~(\ref{equ:A4}) and 
\begin{eqnarray}
\nonumber
 \Delta C(\tilde{P},\tilde{q},\theta_q,\theta_{q'}) 
    & = & \frac{F_D(p(\tilde{P},\tilde{q}^2,\cos\theta_{q'}))}
               {p(\tilde{P},\tilde{q}^2,\cos\theta_{q'})}
         +\frac{F_D(p(\tilde{P},\tilde{q}^2,-\cos\theta_{q'}))}
               {p(\tilde{P},\tilde{q}^2,-\cos\theta_{q'})} \\
 \label{equ:B10}
 & & \hspace{1cm} -\, 
        \frac{F_D(p(\tilde{P},\tilde{q}^2,\cos\theta_q))}
             {p(\tilde{P},\tilde{q}^2,\cos\theta_q)}
       -\frac{F_D(p(\tilde{P},\tilde{q}^2,-\cos\theta_q))}
             {p(\tilde{P},\tilde{q}^2,-\cos\theta_q)} \,, \\
 \nonumber
 C'(\tilde{P},\tilde{q},\theta_{q'}) 
     & = & \frac{p(\tilde{P},\tilde{q}^2,\cos\theta_{q'})
                -[1+2(p(\tilde{P},\tilde{q}^2,\cos\theta_{q'}))^2]
                  F_D(p(\tilde{P},\tilde{q}^2,\cos\theta_{q'}))}
                {2(p(\tilde{P},\tilde{q}^2,\cos\theta_{q'}))^3} \\
 \nonumber
 & & \hspace{1cm} 
     \times \left(2\tilde{q}+\tilde{P}\cos\theta_{q'}\right) \\
 \nonumber
 & & +\,  \frac{p(\tilde{P},\tilde{q}^2,-\cos\theta_{q'})
                -[1+2(p(\tilde{P},\tilde{q}^2,-\cos\theta_{q'}))^2]
                  F_D(p(\tilde{P},\tilde{q}^2,-\cos\theta_{q'}))}
               {2(p(\tilde{P},\tilde{q}^2,-\cos\theta_{q'}))^3} \\
 \label{equ:B11}
 & & \hspace{1cm} 
   \times \left(2\tilde{q}-\tilde{P}\cos\theta_{q'}\right) \,.
\end{eqnarray}
By evaluating the above integrals numerically, we find 
$\mathcal{I}_1\approx -0.02194$, $\mathcal{I}_2\approx 0.23899$, 
$\mathcal{I}_3\approx -0.00059$ and $\mathcal{I}_4\approx -0.18650$. In $c_1$, 
the first and the third term are both large compared to the others, but of opposite sign. 
It is interesting to note that these together with the fourth and fifth term basically 
cancel each other, leaving the $z(\lambda/a)$-dependence of $(\eta-\eta_\infty)/\eta_\infty$ 
determined by the second term in $c_1$ which is related to ${\rm Re}\,\delta\Pi$ in the 
in-medium cross section. 


\end{document}